\documentclass[manuscript,screen,acmsmall]{acmart}
\usepackage{booktabs}
\usepackage{hyperref}
\AtBeginDocument{%
  }

\setcopyright{acmlicensed}
\copyrightyear{2018}
\acmYear{2018}
\acmDOI{XXXXXXX.XXXXXXX}
\acmConference[Conference acronym 'XX]{Make sure to enter the correct
  conference title from your rights confirmation email}{June 03--05,
  2018}{Woodstock, NY}
\acmISBN{978-1-4503-XXXX-X/2018/06}

\begin{document}

\title{Computer-Aided Tagging on Wikimedia Commons: Designing for Human–AI Collaboration in Open Knowledge Work}

\author{Yihan Yu}
\affiliation{ Human Centered Design \& Engineering 
  \institution{University of Washington}
  \streetaddress{3960 Benton Ln NE}
  \city{Seattle}
  \state{Washington}
  \country{USA}
  \postcode{98195}}
\email{yyu2016@uw.edu}

\author{David W. McDonald}\orcid{0000-0001-5882-828X}
\affiliation{ Human Centered Design \& Engineering 
  \institution{University of Washington}
  \streetaddress{3960 Benton Ln NE}
  \city{Seattle}
  \state{Washington}
  \country{USA}
  \postcode{98195}}
\email{dwmc@uw.edu}

\begin{abstract}
This study investigates Wikimedia Commons contributors' lived experiences with the Computer-Aided Tagging (CAT) tool, an AI-assisted image tagging system designed to improve Commons' discoverability, searchability, accessibility, and multilingual support. Using a qualitative analysis of 595 CAT-related community comments from 11 wiki pages and 16 in-depth interviews, we identify seven key issues that contributed to CAT’s mixed reception and eventual deactivation. We also offer community-informed suggestions for improving the tool. We reflect on the implications for designing human–AI collaboration on Commons and for developing AI-assisted tools that support open knowledge work. This work contributes to HCI and CSCW research by extending the understanding of human–AI collaboration beyond Anglophone, text-centric, corporate platforms.
\end{abstract}

\maketitle

\section{Introduction}

Wikimedia Commons is one of the world’s largest repositories of free-to-use images and media files. It supports over 300 Wikipedia language editions and a global community of contributors and users. Commons has relied on a hierarchical category system as its primary infrastructure for organizing media. This system functions as a decentralized, volunteer-driven social tagging mechanism that allows contributors to manually annotate files using nested category tags. While categories support browsing, keyword-based search, and manual organization, Commons’ rapid growth has revealed persistent limitations in discoverability, semantic search, and multilingual access, largely because file metadata remains predominantly unstructured and English-dominant in practice.

To address these limitations, the Wikimedia Foundation launched the Structured Data on Commons (SDC) project. SDC moves Commons metadata from unstructured plaintext into a structured, ontology-linked property–value statement model. It connects file annotations to Wikidata’s shared knowledge graph so that metadata can resolve to a globally consistent, community-maintained ontology. To support community adoption of SDC, the Foundation later introduced the Computer-Aided Tagging (CAT) tool. CAT was designed as a human-in-the-loop image-tagging assistant. It generates AI-suggested labels using the Google Cloud Vision API, maps those labels to language-agnostic Wikidata items, and presents them to Commons contributors for human verification. Contributors can review each suggestion and selectively publish confirmed labels as SDC Depicts statements, which are structured semantic claims that assert what is visually represented in a media file and link files to shared concepts defined in Wikidata. In doing so, CAT aimed to support semantic media search across languages and to make multimedia files more discoverable and reusable across Wikimedia projects.

Prior research in HCI/CSCW has investigated human–AI collaboration on commercial, corporate-controlled platforms such as Facebook \cite{grandinetti2023examining,kuo2023unsung}, Twitter \cite{wischnewski2024agree}, and TikTok \cite{kang2022ai,wang2024reelframer,grandinetti2023examining}. In the context of Wikimedia projects, prior research has primarily focused on AI/ML tools deployed in English Wikipedia that support indirect tasks, such as detecting vandalism \cite{adler2011wikipedia}, evaluating contributions or article quality \cite{halfaker2020ores}, and suggesting tasks \cite{cosley2007suggestbot}. However, little is known about how human contributors and AI collaborate in a multilingual, community-governed platform with a focus on direct content creation rather than indirect tasks. 

In this study, we address this gap by presenting one of the first empirical accounts of the full lifecycle of a human–AI collaboration deployed within a volunteer-governed open knowledge ecosystem, where semantic tagging is central to direct knowledge creation. By studying CAT’s mixed reception and eventual deactivation as a consequential failed AI deployment, we treat failure as an empirical resource to reveal broader, generalizable CSCW insights into how legacy socio-technical infrastructures, community values, goals, and practices, participatory design approaches, and consensus-driven governance mechanisms interact with AI pipelines to shape collaboration breakdowns and possibilities for redesign in open knowledge ecosystems. Our findings generate community-informed design implications for future AI-assisted tools that seek to support SDC.

In the following sections, we first introduce Wikimedia Commons, its category system, SDC, the CAT tool, and tagging practices on Commons. We then review related literature on AI/ML-assisted tools in Wikimedia projects and describe our methods, which include a qualitative analysis of Wiki discussions and participant interviews. The paper reports seven key issues that contributed to CAT’s mixed reception and eventual deactivation, along with community-informed suggestions for improving tool design and adoption. We conclude by discussing implications for human–AI collaboration on Commons and for AI-assisted tools supporting open knowledge work.

\section{Background and Literature Review}

In this section, we introduce Wikimedia Commons, the Commons category system, SDC, and the CAT tool. We then define tagging on Commons and situate our research within the existing literature on the design of AI/ML-assisted tools in Wikimedia projects.

\subsection{Wikimedia Commons}

Wikimedia Commons\footnote{\url{https://commons.wikimedia.org/wiki/Commons:Welcome}} is one of the world’s largest online repositories of freely usable multimedia files, including images, audio, and video. Its content is contributed and curated by a global community of volunteers. According to recent statistics\footnote{\url{https://commons.wikimedia.org/wiki/Special:Statistics}}, the platform hosts over 118 million files contributed by approximately 13 million users. Commons supports more than 300 Wikipedia language editions as well as other Wikimedia projects such as Wikivoyage, Wikispecies, and Wikiversity, while also serving the broader internet public.

Despite its scale and importance, Commons remains relatively understudied and underappreciated in academic research \cite{menking2020image,yu2022unpacking}, particularly in contrast to its sister project, Wikipedia. While both are examples of open knowledge work, prior work \cite{yu2023you,yu2022unpacking} highlights key differences between the two. Wikipedia focuses on collaborative text production for an online encyclopedia, whereas Commons centers on the collection and curation of free-to-use multimedia. Although both platforms are built on the MediaWiki software, their functionalities, symbolic roles (reference vs. collection), and governance structures position them as fundamentally distinct socio-technical platforms, in line with Gillespie’s extended definitions of platform \cite{gillespie2010politics}.

Overall, while Commons plays a critical infrastructural and cultural role within the Wikimedia ecosystem and the broader open knowledge landscape, it is uniquely characterized as a collaboration-for-reuse platform rather than a co-authoring platform \cite{yu2023you,yu2022unpacking}. Unlike text-centric peer production systems such as Wikipedia, which converge on a single, relatively stable knowledge artifact, Commons supports ongoing, distributed collaboration in which contributors share, interpret, curate, and validate visual materials within diverse linguistic and cultural contexts. Its core coordination work centers on iterative negotiation over metadata, categorization, licensing, and cross-project image reuse, rather than the joint authorship of a single fixed object. Commons contributors do not co-produce a singular canonical artifact. Instead, they collectively construct and maintain evolving descriptive infrastructures that enable shared image sensemaking and support downstream reuse at global scale. These distinctive activities and collaboration dynamics make Commons an important and comparatively understudied site of inquiry for CSCW research.

\subsection{Commons Category System}

Wikimedia Commons has relied for over twenty years on a hierarchical category system\footnote{\url{https://commons.wikimedia.org/wiki/Commons:Categories}} as its primary infrastructure for organizing and navigating multimedia content. The system operates as a collaborative, decentralized social tagging mechanism. It allows contributors to manually apply hierarchical category tags to file metadata. Because multiple contributors can annotate the same file from diverse perspectives, the category system has supported distributed sensemaking and the bottom-up construction of a large-scale, broad folksonomy, an evolving, community-driven classification structure that functions without a strictly enforced controlled vocabulary. Despite its longstanding role as the primary navigational tool for search, browsing, and content discovery, Commons’ rapidly expanding scale has exposed fundamental limitations of the category system, particularly in supporting discoverability, searchability, accessibility, and multilingual participation.

First, the system is constrained by the design assumptions of MediaWiki, which was originally developed to support text-centric content rather than image-first repositories. As a result, category tags are human-generated, unstructured plaintext and lack a consistent machine-readable schema \cite{yu2022unpacking}. This constrains advanced retrieval, limits semantic interoperability, and reduces the platform’s ability to integrate metadata with structured knowledge bases such as Wikidata or external ontologies.

Second, although the Commons community has collaboratively built and maintained this categorization system, it suffers from challenges typical of user-generated image tags \cite{wang2012assistive}: they are often noisy, inconsistent, incomplete, or overly subjective. While such tags can aid organization and sensemaking, they also introduce fuzziness and ambiguity, as contributors choose tags based on personal preferences, tendencies, and beliefs \cite{thornton2012tagging,sen2006tagging}. As a result, many tags are meaningful only to their original contributors, which reduces the overall precision of the collaborative tagging system.

Third, Commons’ dependence on manual tagging has left a substantial portion of files uncategorized due to the ongoing time and labor required for human annotation \cite{yu2022unpacking, wang2012assistive}. Unlike text, images are not self-describing, meaning files without category tags are effectively undiscoverable through the primary navigation system. In practice, this has created a long tail of archived images that are neither findable nor browsable, which reduces both archival integrity and reuse potential.

Finally, despite Commons’ commitment to multilingual inclusivity, English remains dominant in category tags \cite{yu2022unpacking}. This language imbalance limits accessibility for non-English-speaking users, who face significant challenges when contributing to, searching for, and discovering relevant content. These linguistic barriers exacerbate broader inequities in participation and access, both within Commons and across the wider Wiki ecosystem \cite{miquel2018wikipedia}.

\subsection{Structured Data on Commons and Computer-Aided Tagging}

To address the limitations of the category system outlined above and improve discoverability, searchability, accessibility, and multilingual participation on Wikimedia Commons, the Wikimedia Foundation launched the Structured Data on Commons (SDC) project\footnote{\url{https://commons.wikimedia.org/wiki/Commons:Structured_data}}. SDC introduces a metadata model for media files that is both human-interpretable, so contributors can read and edit it directly, and machine-actionable, so the data follows a predictable, standardized structure that software systems can parse uniformly at scale.

Before SDC, Commons files were primarily annotated using unstructured plaintext metadata, including filenames, free-form descriptions, and hierarchical category tags. These annotations are typically written in a single language (most often English), lack semantic typing, and do not conform to a shared data schema. As a result, software systems cannot reliably interpret the meaning of category tags or link them to common concepts. This limits semantic retrieval, multilingual rendering, and interoperability with structured databases. SDC replaces this model with a property–value statement framework grounded in Wikidata\footnote{\url{https://www.wikidata.org/wiki/Wikidata:Main_Page}}, the Wikimedia ecosystem’s central structured knowledge graph, which stores concepts and real-world entities as uniquely identified items and links them through a shared, community-governed ontology. In this framework, each metadata claim is expressed as a semantically defined property (e.g., Depicts, Creator, License) paired with a value that references a globally unique Wikidata identifier (Q-ID). Because both properties and values resolve to Wikidata’s ontology, Commons file metadata can interoperate with a broader, interconnected semantic ecosystem that spans Wikimedia projects and external databases. Wikidata identifiers are language-independent. This allows the same SDC statements to be rendered dynamically across languages through interface localization, without requiring contributors to manually duplicate or translate tags.

\begin{figure}[ht]
  \caption{A side-by-side comparison of a file’s traditional, unstructured File information view (right) and its structured data view (left).}
  \label{fig:sdc}
  \centering
  \includegraphics[scale= .5]{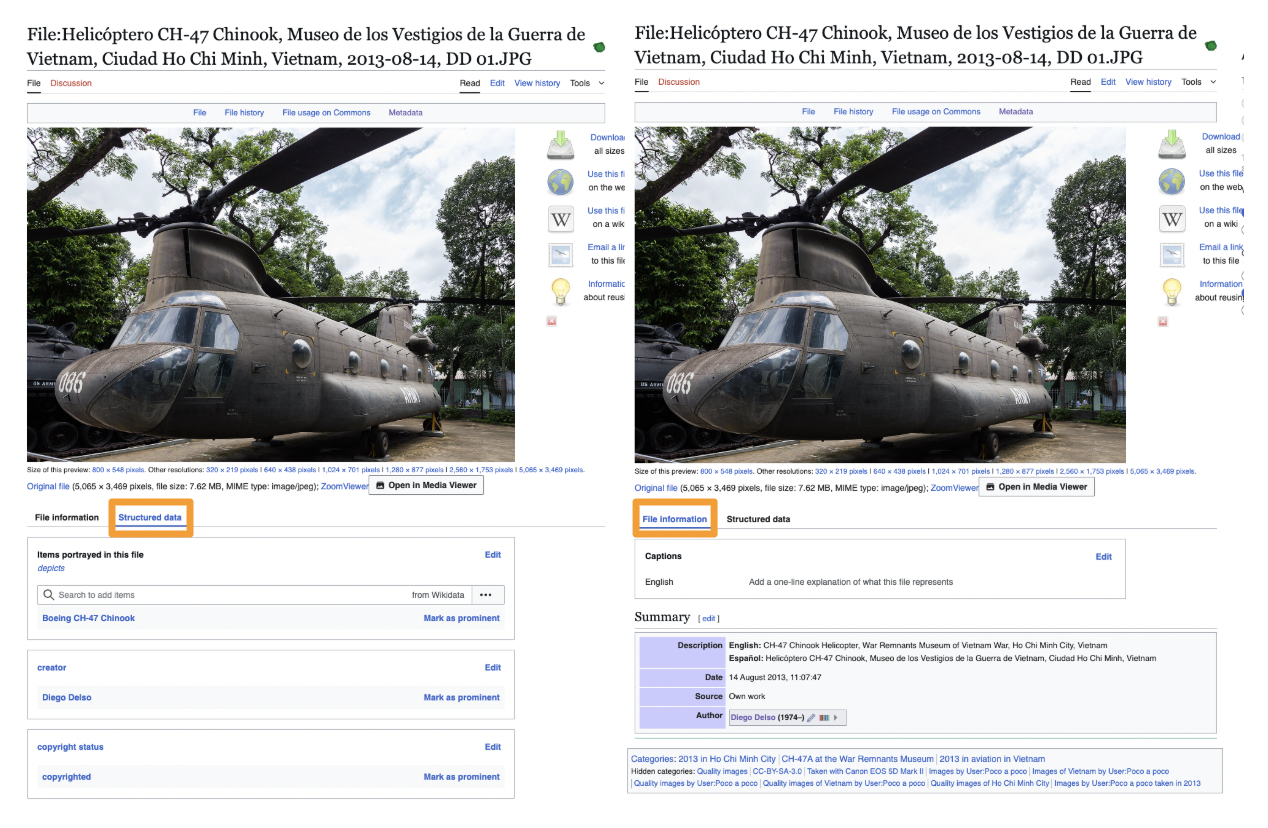}
\end{figure}

Every media file page on Wikimedia Commons now includes both a File information tab, which contains traditional, unstructured metadata such as wikitext descriptions and contributor-created categories, and a Structured data tab, which displays SDC statements encoded in a structured, ontology-linked format. Under the Structured data tab, contributors can assert what is visually represented in a file using Depicts (P180) statements and can also add other structured claims, such as license (P275), creator (P170), copyright status (P6216), and community-maintained quality or compliance assessments. Figure \ref{fig:sdc} provides a side-by-side comparison of a file’s traditional, unstructured File information view and its structured SDC statement view. Illustrative examples of SDC statements shown under the Structured data tab include:

\begin{itemize}
    \item Depicts (P180): Boeing CH-47 Chinook (Q209613)
    \item Copyright status (P6216): Copyrighted (Q50423863)
    \item Creator (P170): Diego Delso (Q28147777)
\end{itemize}

To support community adoption of SDC, the Wikimedia Foundation later introduced the Computer-Aided Tagging (CAT) tool.\footnote{\url{https://commons.wikimedia.org/wiki/Commons:Structured_data/Computer-aided_tagging}} CAT was designed as a human-in-the-loop image tagging assistant that helps contributors identify and apply a single structured metadata field in SDC: Depicts (P180) statements.\footnote{\url{https://commons.wikimedia.org/wiki/Commons:Depicts}} Depicts statements specify what is visually represented in a media file.

The CAT tool uses the Google Cloud Vision API\footnote{\url{https://cloud.google.com/vision}}, a commercial image-recognition service provided by Google. In this study, we and our participants describe this system as \textit{generic} because it is a general-purpose model not designed or customized for Wikimedia Commons. While we cannot rule out that some Commons files may be included in Google’s broader training data, the model is not specifically trained for computer-aided tagging on Commons, does not incorporate feedback from Commons contributors, and is not developed or maintained by Wikimedia Foundation researchers. 

\begin{figure}[ht]
  \caption{The CAT interface displaying unverified suggested tags for an example image file.}
  \label{fig:interface}
  \centering
  \includegraphics[scale= .6]{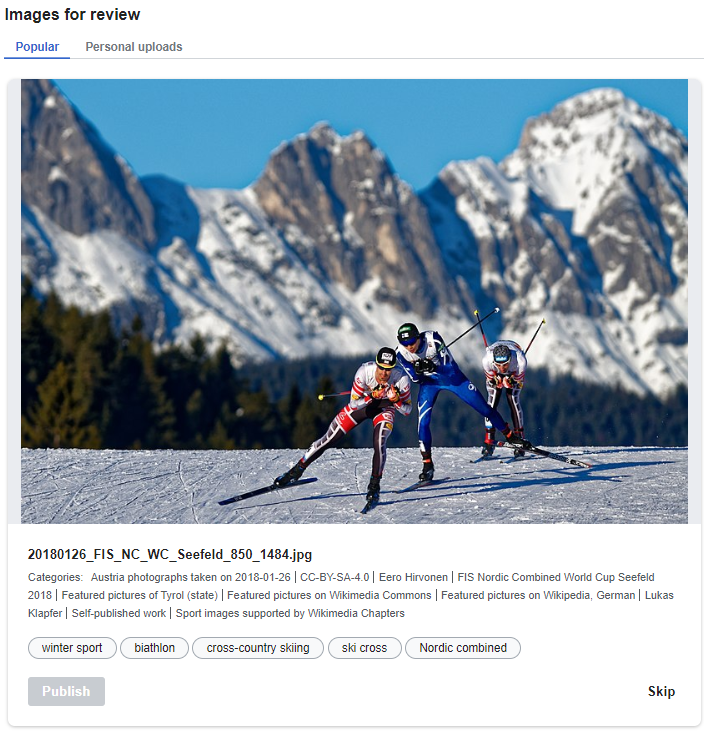}
\end{figure}

After an image is uploaded to Commons, the API analyzes the visual content alone, without accessing or incorporating any existing unstructured file metadata, and returns candidate labels. Each label is associated with a Google Knowledge Graph concept ID, which CAT maps to a Wikidata Q-ID using the Freebase–Wikidata mappings\footnote{\url{https://developers.google.com/freebase/\#freebase-wikidata-mappings}}. The mapped Wikidata entities are surfaced to contributors as suggested tags in the CAT interface. Figure \ref{fig:interface} shows an example of this interface, which displayed unverified suggested tags beneath an uploaded image. Contributors could evaluate, click individual suggestions to accept them, and select “Publish” to save confirmed tags. They could also skip all suggestions. Only human-confirmed tags were saved, and each accepted tag was stored as a Depicts (P180) → Wikidata Item (Q-ID) statement in the file’s structured metadata.

\begin{figure}[ht]
  \caption{The Structured data tab showing Depicts statements (left) alongside traditional category tags stored as plaintext markup (right).}
  \label{fig:depicts}
  \centering
  \includegraphics[scale= .5]{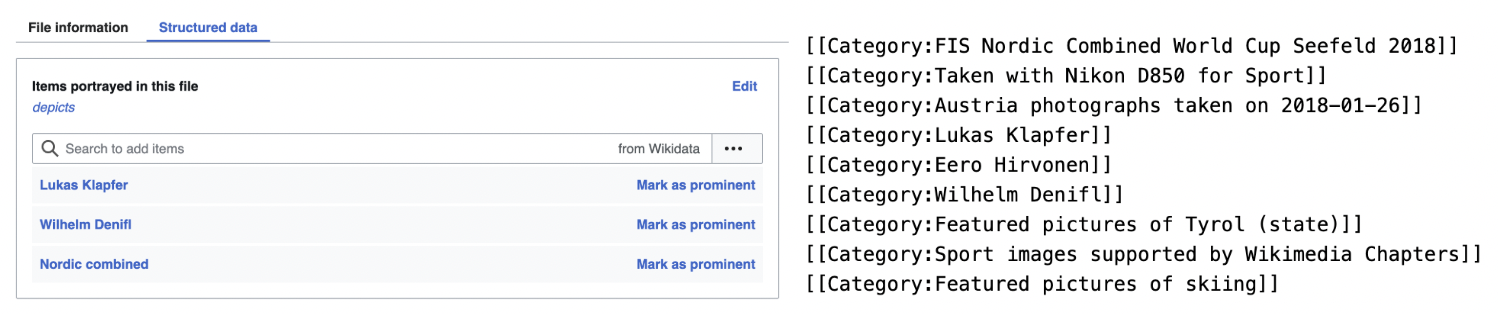}
\end{figure}

Figure \ref{fig:depicts} shows how confirmed Depicts statements (left) appear on the file page of the same example image shown earlier in Figure \ref{fig:interface}, in contrast to that image’s traditional category markup (right), which consists of hierarchical, contributor-generated plaintext tags that are not semantically typed and not linked to a shared ontology. While categories support human browsing and manual organization, Depicts statements enable multilingual semantic search based on linked identifiers and meaning, rather than keyword-based string matching.

The CAT tool was deactivated in September 2023.

\subsection{Defining Tagging on Wikimedia Commons}

In the previous two subsections, we introduced three distinct ways to “tag” an image file on Wikimedia Commons. First, contributors can add category tags—the traditional method of organizing files by placing them into human-defined topic groups. We refer to this process as \textbf{\textit{categorizing}}.

Second, the CAT tool uses the Google Cloud Vision API to automatically label images, generating suggested tags based on visual content analysis. These are machine-generated recommendations for what might be included in a Depicts statement. We refer to this process as \textbf{\textit{suggesting tags}}.

Third, human contributors review the suggested tags via the CAT interface and selectively apply them as Depicts statements—structured, multilingual data that semantically describe the visual elements of an image. We refer to this activity as \textbf{\textit{applying Depicts statements}}. 

Across all three processes, we use the term \textbf{\textit{“tagging” as a verb}} to refer to the general activity of applying metadata to images. What differs is the type of tag being applied and who or what is doing the tagging (a human or a machine). To maintain clarity throughout the paper, we adopt the following terminology:

\begin{itemize}
    \item We use “\textbf{\textit{category tags}}” to refer to traditional tagging in Commons’ category system.
    \item We use “\textbf{\textit{suggested tags}}” to refer to the AI-generated outputs of the CAT tool.
    \item We use “\textbf{\textit{Depicts statements}}” to refer to structured tags applied by human editors using the CAT tool.
\end{itemize}

This distinction allows us to describe tagging as a collaborative process between humans and AI, while remaining precise about the specific forms and sources of tagging involved. We note that quotes from discussion posts and participant interviews use 'tag' and 'tagging' in a much less precise manner.

\subsection{Designing AI/ML-Assisted Tools in Wikimedia Projects}

HCI and CSCW research has examined human–AI collaboration in systems that support moderation, recommendation, and content authoring on corporate-controlled platforms such as Facebook \cite{grandinetti2023examining, kuo2023unsung}, Twitter/X \cite{wischnewski2024agree}, and TikTok \cite{kang2022ai, wang2024reelframer, grandinetti2023examining}. These systems are introduced top-down and optimized for proprietary metrics such as engagement and retention, with AI behaviors governed centrally by platform owners and designers. In contrast, open knowledge work projects are governed by decentralized communities of volunteers \cite{forte2009decentralization}. The platform owners do not make editorial decisions. Rather, decisions about content creation, moderation, and tool deployment are made through open participation, consensus-building, and shared responsibility. These socio-technical differences shape how AI/ML tools are designed, adopted, and evaluated within open knowledge work contexts \cite{halfaker2020ores}.

Nowadays, CSCW research has shown growing interest in developing AI/ML-assisted tools to support open knowledge work. In particular, prior research has designed, deployed, and evaluated tools in English Wikipedia that support indirect tasks \cite{kittur2007he, viegas2007talk}, such as vandalism detection, contribution or article quality assessment, and task suggestion. For example:

\begin{itemize}
\item Quality control tools \cite{adler2011wikipedia} help patrollers evaluate edit quality, and identify and revert harmful edits.
\item Recommendation systems, such as SuggestBot \cite{cosley2007suggestbot}, suggest articles or tasks in need of attention based on an editor’s interests or activity history.
\item Meta-algorithmic systems, such as ORES \cite{halfaker2020ores, smith2020keeping} and LiftWing,\footnote{\url{https://wikitech.wikimedia.org/wiki/Machine_Learning/LiftWing}} represent a deeper integration of Wikimedia's participatory values into AI/ML infrastructure. These tools are not only used to make edit or article quality predictions in support of moderation, they also serve as platforms that enable community members to audit, contest, and adapt models to local needs and values to support transparent and collaborative development.
\end{itemize}

However, these systems have been studied exclusively in the context of Wikipedia, with a focus on indirect work, editorial moderation and coordination, rather than on direct content creation. Current research therefore overlooks how AI/ML-assisted tools support the core goal of open knowledge work, which is the direct, community-governed creation of new shared knowledge \cite{kittur2007he, viegas2007talk}.

\begin{figure}[ht]
  \caption{Screenshots from an animation demonstrating multilingual structured data statements. The example shows the structured data of Würfelzucker (2018), CC BY-SA 4.0, by Dietmar Rabich.}
  \label{fig:m}
  \centering
  \includegraphics[scale= .5]{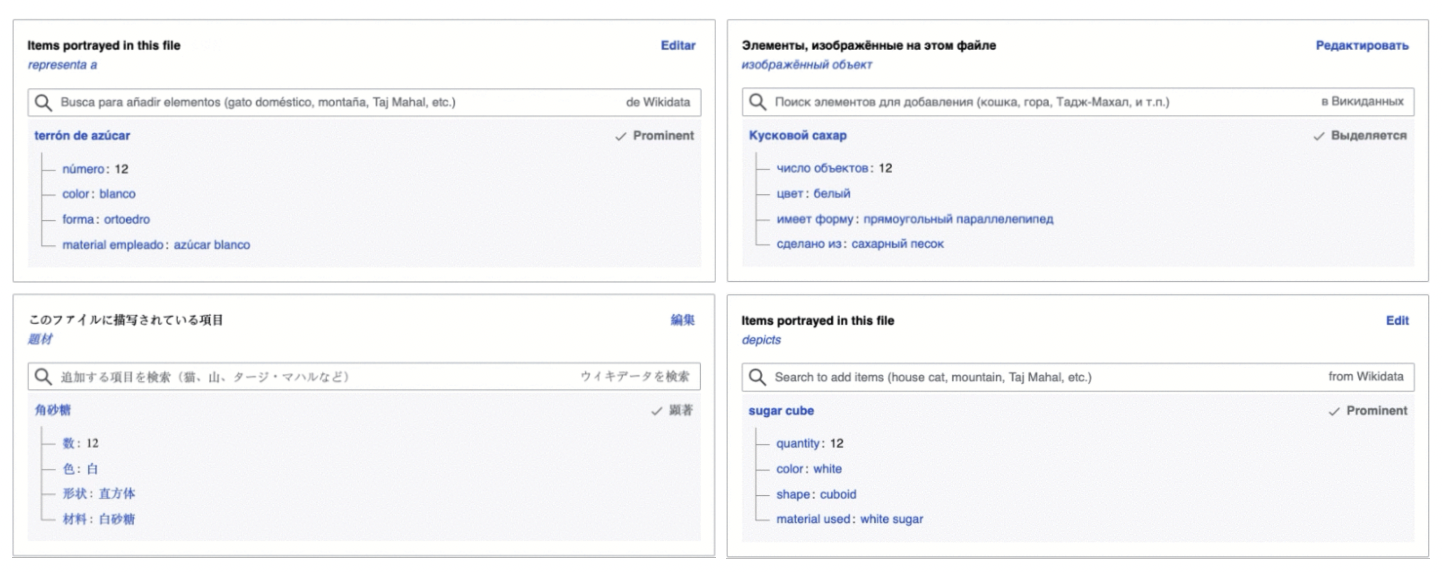}
\end{figure}

Our study shifts focus to a different Wikimedia project: Wikimedia Commons, a multilingual, multimedia-oriented platform that communities across more than 300 language editions of Wikipedia contribute to and use. We investigate Computer-Aided Tagging (CAT), an AI-assisted tool that supports the addition of structured, machine-readable Depicts statements to images. Unlike prior tools, CAT has multilingual affordances. Its design enables contributors to interact with AI-suggested tags in their selected Commons interface language. Because each suggestion is linked to a Wikidata entity, both the AI-suggested tags and the CAT display are automatically translated into multiple languages. In this paper, we use the term \textit{non-Anglophone} to describe these multilingual functionalities, which allow editors who do not primarily work in English to contribute effectively, a capability not supported by the traditional Commons category system or by other AI/ML tools previously deployed in Wikimedia projects. Figure \ref{fig:m} illustrates this functionality by showing how added Depicts statements are displayed in different languages when the user’s interface language setting is changed.

While applying image tags may be seen as a metadata task in other contexts, the Commons community views adding image tags such as Depicts statements as a form of open knowledge creation \cite{yu2022unpacking}. These tags provide semantic information about what an image contains and why it matters, which enriches Commons as a shared media repository, improves image discovery, and contribute directly to the content layer that supports other Wikimedia projects. Importantly, these annotations do not merely describe what is visible in an image, they embed the image within broader conceptual and cultural frames \cite{yu2023you}. Such editorial decisions shape how knowledge is organized and how users around the world find and interpret visual media. From the community’s perspective, adding image tags is not just a technical task, it is an editorial process that generates, structures, and shares knowledge, which constitutes a form of content creation that Commons is built to provide \cite{yu2022unpacking}.

This context introduces unique challenges for AI-assisted tools. Compared to text-based tasks like evaluating edit quality, labeling visual content is more ambiguous, context-dependent, and subjective \cite{shatford1986analyzing}. A single image may require multiple labels, as it can contain several objects, nested elements, or conceptual hierarchies. Additionally, the interpretation of visual features can vary across cultural contexts \cite{foster2016managing}. Unlike structured data with objective metrics (e.g., word count), image tagging depends heavily on culture, personal judgment, and lived experience—factors that are difficult to encode into consistent labeling rules.

In this study, we investigate how human contributors and AI collaborate in a multilingual, community-governed environment, with a focus on this form of content creation. By examining Commons contributors’ experiences with the CAT tool, and the design considerations needed to support its adoption, interaction, and satisfaction, we expand current HCI and CSCW research by offering insights into AI design in contexts beyond Anglophone, text-centric, corporate platforms.

\section{Methods}

To capture both the breadth of community input and the depth of individual perspectives, we combined qualitative data analysis and interview methods. The study was reviewed by our institution's Institutional Review Board (IRB) and was determined to be exempt. We followed the Wikimedia Foundation's recommendation to create a research project page on Meta-Wiki. This page described our project goals, methods, progress, and potential impacts.

\subsection{Qualitative Analysis of Wiki Discussions Related to the CAT Tool}

We conducted a qualitative analysis of wiki discussions about the CAT tool for a broader understanding of the Commons editing community's needs, experiences, and concerns regarding the tool. We identified 11 wiki pages (Table \ref{tab:wikipages}) where the CAT tool was discussed. These pages contained 595 comments across 172 topics, contributed by 160 unique wiki users. Table \ref{tab:distri} shows the distribution of user engagement in CAT-related discussions. Seven users left more than 10 CAT-related comments, 15 users contributed 5–10 comments, 59 users made 2–4 comments, and the majority—79 users—left only one comment. 

\newcommand{\vpumpURL}{https://commons.wikimedia.org/w/index.php?title=Commons:Village_pump\&oldid=399548798\#Depicts}
\begin{table}[ht]
\caption{Analyzed Wiki pages}
\label{tab:wikipages}
\centering
\resizebox{\columnwidth}{!}{%
\begin{tabular}{llll}
\toprule 
\bf Page & \bf {\#topics} & \bf {\#comments} & \bf {Page Type}\\
\midrule
\href{https://commons.wikimedia.org/wiki/Commons_talk:Structured_data/Computer-aided_tagging}{Page 1} & 6 & 37 & Commons talk page\\
\href{https://commons.wikimedia.org/wiki/Commons_talk:Structured_data/Computer-aided_tagging/Archive_2019}{Page 2} & 24 & 58 & Commons talk page\\
\href{https://commons.wikimedia.org/wiki/Commons_talk:Structured_data/Computer-aided_tagging/Archive_2020}{Page 3} & 111 & 349 & Commons talk page\\
\href{https://commons.wikimedia.org/wiki/Commons_talk:Structured_data/Computer-aided_tagging/Archive_2021}{Page 4} & 12 & 15 & Commons talk page\\
\href{https://commons.wikimedia.org/wiki/Commons_talk:Structured_data/Computer-aided_tagging/Archive_2022}{Page 5} & 9 & 11 & Commons talk page\\
\href{https://commons.wikimedia.org/wiki/Commons_talk:Structured_data/Computer-aided_tagging/Archive_2023}{Page 6} & 5 & 10 & Commons talk page\\
\href{https://phabricator.wikimedia.org/T339902}{Page 7} & 1 & 7 & Phabricator task page\\
\href{\vpumpURL}{Page 8} & 1 & 15 & Commons village pump discussion\\
\href{https://commons.wikimedia.org/wiki/Commons:Village_pump/Proposals/Archive/2020/04}{Page 9} & 1 & 5 & Commons village pump proposal\\
\href{https://www.wikidata.org/wiki/Wikidata:Property_proposal/tag}{Page 10} & 1 & 8 & Wikidata property proposal\\
\href{https://commons.wikimedia.org/wiki/Commons:Village_pump/Archive/2020/02}{Page 11} & 1 & 80 & Commons village pump discussion\\
\midrule
\bf Total & \bf 172 topics & \bf 595 comments & \bf 11 pages\\
\bottomrule
\end{tabular}%
}
\end{table}

The first author manually copied and pasted all 595 comments, along with the corresponding user names and links to the user pages we identified either on Wikimedia Commons or Meta-Wiki, into a spreadsheet for analysis. We then conducted open coding of the comments using a thematic analysis approach \cite{braun2006using}. This process identified seven emergent themes regarding Commons editors’ experiences with the deactivated CAT tool:

\begin{itemize}
    \item Goals of the Structured Data on Commons project,
    \item Evaluations of the quality of suggested tags,
    \item Definitions of Depicts statements,
    \item Differences between tags and Depicts statements,
    \item Existing infrastructure (image titles, categories, descriptions),
    \item Documentation and instructions for the tool, and
    \item User interface issues, including difficulties skipping images, overwhelming notifications, editing suggested tags, adding additional tags, error messages, and waiting periods.
\end{itemize}

Findings of this qualitative analysis informed both the recruitment of participants and the development of our interview protocol.

\begin{table}[ht]
  \caption{Distribution of User Engagement}
  \label{tab:distri}
  \centering
  \begin{tabular}{l l}
    \toprule
    \textbf{\#Comments} & \textbf{\#Users} \\
    \midrule
    More than 10 & 7 \\
    5–10 & 15 \\
    2–4 & 59 \\
    1 & 79 \\
    \bottomrule
  \end{tabular}
\end{table}

\subsection{Interviews}

A deeper understanding of Commons editors’ experiences with the deactivated CAT tool and the issues identified in our qualitative coding of CAT-related discussions came from interviews with Commons editors who had participated in those discussions.

\subsubsection{Recruitment}

We aimed to recruit all 160 unique users that we identified who had participated in CAT-related discussions. However, we did not or were unable to send invitations to 41 users due to various reasons:

\begin{itemize}
    \item Six user pages appeared to belong to Wikimedia Foundation researchers responsible for designing and/or developing the CAT tool.
    \item Twenty-six users did not have a user page on Commons or Meta Wiki.
    \item One user page was attributed to a deceased Wikipedian (WP:RIP).
    \item One user was banned from editing all Wiki projects indefinitely.
    \item One user was marked as retired from editing Wiki projects (WP:RETIRE).
    \item Six users had user pages on Commons or Meta-Wiki but did not enable the “Email this User” function.
\end{itemize}

We contacted the remaining 119 Commons users using the built-in “Email this User” function on Commons and/or Meta-Wiki. Each editor received a personalized message including their username, the purpose of the study, a link to our Meta-Wiki study page, the wiki page where their participation in CAT-related discussions was identified, and an invitation to participate in an interview. We also noted that participation in the study was voluntary and offered a \$25 Amazon gift card as a token of appreciation for their time, effort, and expertise.

Of the 119 editors contacted, 29 responded, and 16 completed an interview with us. Table \ref{tab:participant} shows the demographic information of these 16 interview participants.

\begin{table}[ht]
\caption{Demographic information of interview participants}
\label{tab:participant} 
\centering
\resizebox{\columnwidth}{!}{%
\begin{tabular}{lllll}
\toprule
\textbf{Pseudonym} & \textbf{Interview Method} & \textbf{Commons Edit Count Cohort} & \textbf{Wikipedia Language Edition} & \textbf{Wikidata Editor} \\
\midrule
Mark & Video Conference & 600,001+          & English \& German       & Yes \\
Ian & Video Conference & 0 – 25,000        & English             & No  \\
Victor & Video Conference & 0 – 25,000        & English, Russian \& Belarusian   & Yes \\
Kevin & Video Conference & 25,001 – 100,000  & English             & Yes \\
James & Online Chat      & 0 – 25,000        & English             & No  \\
Vin & Video Conference & 100,001 – 300,000 & English \& French       & Yes \\
Jack & Video Conference & 300,001 – 600,000 & English             & Yes \\
Joel & Video Conference & 300,001 – 600,000 & English \& Polish       & Yes \\
Mira & Video Conference & 25,001 – 100,000  & English             & Yes \\
Harris & Video Conference & 0 – 25,000        & English \& German       & Yes \\
Cody & Online Chat      & 300,001 – 600,000 & English, Spanish \& French   & Yes \\
Sam & Video Conference & 600,001+          & English             & Yes \\
Ron & Phone Call       & 25,001 – 100,000  & English \& German \& French & Yes \\
Kirk & Video Conference & 100,001 – 300,000 & English             & Yes \\
Paul & Video Conference & 300,001 – 600,000 & English             & Yes \\
Simon & Video Conference & 25,001 – 100,000  & English \& Japanese       & Yes \\
\bottomrule
\end{tabular}%
}
\end{table}

\subsubsection{Data Collection and Analysis}

Guided by the findings from the qualitative coding, we developed an interview protocol (Appendix \ref{app:script}) consisting of an opening script, four interview phases, and a closing script.

In the opening script, we introduce ourselves, explain the purpose of the interview, and outline how we handle and maintain the confidentiality of participants’ data. We also request consent to audio record the session. Before participants provide consent, we clarify that despite our efforts to anonymize data, Wikipedians may sometimes identify individuals or situations from the discussion. This ensures that participants are aware of any residual risks to confidentiality.

In the first phase of the interview, we ask introductory questions to understand participants’ engagement with Commons. These questions include how long the participant has been editing Commons and what types of work they typically do on the platform. We also ask participants to explain, in their own words, what they understand to be the goals of the Structured Data on Commons project. These questions help establish rapport and gather foundational information about the participants’ perspectives and experiences.

The second phase of the interview focuses on a discussion of sample images that participants recently uploaded to Commons. We show them three images they recently contributed and, for each image, ask them to describe the story behind the contribution, what the image depicts, and how they would tag the image themselves. Following this, we provide a list of suggested tags generated by the Google Cloud Vision API for the image, matched to Wikidata items, to replicate the type of suggestions made by the deactivated CAT tool. We then ask participants to reflect on these suggested tags, including whether they agree with any of the suggested tags, whether they would accept and add any of the suggested tags to the Depicts statements, how the suggested tags could improve the discoverability of the image on Commons, and what metadata they find important for tagging such images.

The third phase of the interview shifts to reflections on the CAT tool itself. We begin by asking participants what they remember about the tool and how it impacted their work on Commons. We then ask them to explain their understanding of the role of the category system on Commons, the purpose of the Depicts statements, and the differences between these two elements. We also ask them to reflect on the goals of the CAT tool on Commons and whether they think the tool achieved those goals.

In the final phase, we invite participants to share their thoughts on broader issues, such as the risks associated with using AI/ML for tagging images and ways to mitigate these risks on Commons.

In the closing script, we invite participants to share any additional questions or comments. We also acknowledge their time and effort by offering a \$25 Amazon gift card as a token of appreciation and confirm their preferred email address for sending the gift card.

We conducted all 16 semi-structured interviews in October, November, and December 2024. Fourteen of these were conducted in English using participants’ preferred teleconferencing applications (Zoom or Google Meet) or via phone calls. During 13 of the teleconference interviews, we shared our screens to display and discuss participants’ example contributions and the suggested tags. For the phone interview, we sent the participant a list of relevant links in advance and asked them to open the materials on their own devices prior to the session. The duration of these interviews ranged from 45 to 90 minutes. Additionally, two interviews were conducted through online chat. In these cases, the researchers emailed the interview questions to participants, waited for their responses, and subsequently asked follow-up questions. We believe this asynchronous approach provided flexibility for participants who preferred written communication and/or translator. 

These interviews resulted in a dataset of 16 transcripts. The first author transcribed all interviews and documented relevant pages discussed during the sessions—such as user pages, user contributions pages, and Commons talk pages—to support data triangulation. The first author open-coded the transcripts using a thematic analysis approach \cite{braun2006using} and recorded analytical memos. We iterated on the themes and memos and chose to report only the emergent themes related to issues that inhibit contributors' collaboration with the CAT tool. 

\section{Findings}

In this section, we report our findings on seven key issues that contributed to CAT’s mixed reception and eventual deactivation: (1) misalignment in the perspectives of Structured Data on Commons, (2) unclear definitions of the Depicts statement, (3) difficulties applying Depicts through CAT, (4) lack of integration between categories and CAT, (5) an ill-specified AI/ML task, (6) limited support for collaborative evaluation, and (7) a disconnect between CAT and Commons’ search functionality.

\subsection{Misalignment in the Perspectives of Structured Data on Commons}

The Structured Data on Commons project was introduced by the WMF as an innovative initiative to improve the searchability and accessibility of Commons content. However, feedback from participants within the editing community reveals significant dissatisfaction with how the project was executed. While the idea behind Structured Data on Commons was seen as promising and valuable, its implementation was viewed as flawed—both technically and, more critically, socially. One of the main issues identified was that the project was not sufficiently evangelized to the editing community, including long-time contributors. As a result, many were either unaware of its purpose or unclear about how to contribute, which led to gaps in participation. As Vin explains:

\begin{quote}
\textit{"I think the thing is...structured Commons was a very nice idea at the beginning, but it was not exactly well implemented, a bit technically, but mostly socially. It was not introduced well enough to people, so a lot of people, including long timers on commons, don't know it well."} – Vin
\end{quote}

Participants also highlighted a significant misalignment between the Wikimedia Foundation’s vision for Structured Commons and the perspectives of the editing community. Participants expressed frustration that the Foundation’s goals of the project often diverged from the practical realities and needs of those actively contributing to Commons. As Sam noted, the gap in understanding between the two groups led to significant conflict, which continues to affect the project's progress:

\begin{quote}
\textit{"The understanding of what the problem was at the foundation diverged a lot from the understanding of the problem in the editing community, and that divergence led to conflict that we're still recovering from."} – Sam
\end{quote}

The central point of misalignment was the project's perceived focus shift towards generic image tagging, searchability, and discovery—goals that many participants feared would undermine Commons’ educational purpose. Commons has long been seen as a unique resource for educational media, which offers images that are not only freely available but also rich in context and specific information. Participants argued that if Commons were to shift towards prioritizing broad and generic search functionality, it could lose the educational value that has set it apart from other platforms. As Sam stated:

\begin{quote}
\textit{"That's not why you search on Commons. That's why you search on google search right?...Like other image search engines should be the ones assigning these kinds of automatic tags for discovery, because the point of Commons discovery is not generic search. There was a real clear purpose for the editing community and the function that common serves, Right?...The point of Commons is to create an educational media repository, not to create a stock photo repository...If we make it too generic, you ruin the educational value of the media, and you turn it into another stock photo thing...and the reason every LLM is using Wikipedia to train for their models is because we do weird, specific things, like we want the weird bits of humanity, super specific, right? Like that is the reason why our data is helpful for everyone else, and that's the distinguishing feature of us. Is that weird specificity, right?"} - Sam
\end{quote}

Further emphasizing this point, another participant, Joel argued,

\begin{quote}
\textit{"That's kind of...a philosophical discussion of what is the use of the pic, and that for me, the main use is to find a limited number of images that are best at depicting something I don't have many images for. If I'm looking for photographs of unknown faces, you know, I'm sure I can easily find photographs of faces of unknown people. It's not going to be very hard to find those, but what is hard is to find, let's say, a specific image of some moment."} - Joel
\end{quote}

This critique points to a deeper philosophical debate about the role of Commons in the digital ecosystem compared to other platforms like Google as a media search tool or Flickr. While these larger platforms are designed to handle broad, general search queries, Commons’ role, as seen by its editing community, is more specialized. It serves the needs of researchers, educators, and those seeking highly specific, context-rich, and educationally significant multimedia content.

To address this perceived misalignment, participants suggested the need for a clear, evidence-based understanding of who the users of Commons are, why they use it, and how they engage with its content. This can be achieved through user research, data analysis (traffic \& page view analysis, search term tracking \& visualization), and more direct engagement with the Commons community, as Sam pointed out that similar studies had been conducted on Wikipedia,

\begin{quote}
\textit{"We didn't even have evidence...having been inside the foundation, we had no evidence that what our audience ever like, and we still don't. And we're actually discovering, like, on the Wikipedia audience, like our main users are students, retirees, and experts, and, you know, people with deep knowledge desires, not superficial knowledge desires, right?"} – Sam
\end{quote}

Once there is a clearer understanding of the user base and their needs, the Foundation should facilitate a community-wide discussion to revisit and clarify the purpose, functionality, and implementation of Structured Data on Commons. This conversation should aim to align the project with the broader role of Commons as a platform that supports its diverse contributor and user communities.

\subsection{Unclear Definitions of the Depicts Statement}

The CAT tool adds human-approved or edited tags, suggested by AI, to a structured data property—the Depicts statement on Commons. However, participants reported a significant lack of clarity around what Depicts means in this context. This ambiguity has led to confusion and inconsistent use of the tool. As Vin, an experienced Commons contributor, explained:

\begin{quote}
\textit{“It's still kind of new, and you're like, still lost so, there's no clear rule or guidelines. So that's problematic if you don't have a good start point, you go nowhere. And then the technique itself... The tool or the technique you use is following the rules that are mostly non existent. So where's that? Is the tool good compared to what we expect from it? … It's not the problem of the tool. It's not the tool's fault if we don't have a key rule on what we expect.” }– Vin
\end{quote}

Vin’s comment points to an important issue: the absence of well-defined rules and guidelines for how Depicts should be understood and applied on Commons. This lack of clarity forces contributors to rely on personal interpretation, which often leads to inconsistent practices. During our interviews, participants frequently expressed uncertainty when asked to help add Depicts statements. Even when working with their own uploads, they struggled to determine the most appropriate or accurate Depicts statements. As Kevin noted:

\begin{quote}
\textit{“Well, I would say it depicts a Putnam's jumping spider, maybe the obvious thing. And you know, I mean, well this also depends on how granular you want to be, which is part of the big controversy about structured data on Commons—how specific should the data be about depiction? You could say, you know, this depicts a spider; you could say it depicts an animal; you could say it depicts a jumping spider; you could say it depicts eyes, legs, and a leaf. But I think there's never been good agreement on how to use the depict statement on Commons.”} – Kevin
\end{quote}

This ambiguity affects contributors at all levels of experience. Even more experienced editors face ongoing uncertainty over how specific or broad their Depicts statements should be. In the absence of shared policies or guidelines, contributors often operate based on informal group norms or personal judgment. As Jack shared:

\begin{quote}
\textit{“I think to this day, there is not a terribly good definition of what the depicts statement is for. I constantly run into issues with people adding a depicts statement to my own photo. I'll have a picture of a tree, and someone will say, 'depicts Seattle.' Well, I'm sorry, but I don't think anyone really thinks a picture of a tree depicts Seattle in a useful way. So, no, I don't think there's consensus. There’s a group that has a consensus among them, but I'd estimate that half the people using the feature aren't part of that group. It's a loose consensus.”} – Jack
\end{quote}

This absence of shared guidelines often leads to inconsistent practices and even results in "edit wars," where editors remove Depicts statements that they personally disagree with. Participants suggested that future CAT tool development should address this foundational confusion about the Depicts statement on Commons. They recommended that the Wikimedia Foundation collaborate with the Commons editing community to create clear and comprehensive policies or guidelines that define the purpose of the Depicts statement and provide practical examples illustrating its application in different contexts. Participants proposed using the existing consensus on category usage as a model for developing a similar framework for Depicts statements. As one experienced editor, Jack, explained:

\begin{quote}
\textit{"With categories, we have a firm consensus we want the most precise, and the only exceptions to that are when there's a really deep reason to put something higher up the hierarchy...And so...I would say with categories, we've got a pretty clear and strong consensus how to do that or not. I do not believe with depicts they have anything close to a consensus of how they should be used in these circumstances."} - Jack
\end{quote}

Jack’s analogy with the category system suggests that a similar example-based tutorial could be created for Depicts statements to guide the community’s practice around structured data. 

\subsection{Challenges in Applying Depicts via CAT}

In addition to definitional ambiguity, users faced challenges in applying Depicts statements effectively when working with the CAT tool. For newer contributors, these challenges were often amplified by the tool’s suggested tags, which were not always accurate or contextually appropriate. As Vin noted, contributors needed to use their own judgment carefully when approving suggested tags:

\begin{quote}
\textit{“What is suggested? Oh, everything is suggested… But maybe you can be more cautious, because it's just suggestion. You don't have to accept everything. It's not always relevant.. If everyone is very thoughtful and thinks about what correct and what's wrong before approving something, then this might be all right...”} – Vin
\end{quote}

However, new contributors frequently lacked the context or training to make such thoughtful decisions. The CAT tool’s gamified design—encouraging high volumes of tagging—further contributed to misuse, as Sam observed:

\begin{quote}
\textit{“They deployed it to a bunch of new editors, which was like, just a bad, bad idea, right? … some users are like, Oh, more tags is better. And it's like, actually, that's not the purpose of depicts, right? More is not necessarily better… You know, it's not all of them, but it's always the ones that are like, I just want more. I want more. And so like the gamification, because it's driving at do more instead of solve a specific problem.”} – Sam
\end{quote}

This lack of clear guidance often led new users to conflate quantity with quality. As a result, many added irrelevant or misleading Depicts statements, assuming they were following best practices. Paul gave an example:

\begin{quote}
\textit{“There was another example where the tag was green, it was a picture of the countryside, and the predominant color was green, and the tool had suggested green, so the user had clicked the button and made it say, depicts green...So for two years, innocent people had been doing bad things because the tool asked them to do it, and they thought it was what was wanted. It wasn't their fault, but it was the fault of the people who were asking them to do that.”} – Paul
\end{quote}

Without a clear definition of Depicts or sufficient support on how to apply CAT suggestions responsibly, contributors, particularly newcomers, often misapplied the feature. These misapplications produced lower quality data and contributed to frustration among more experienced editors, who had to correct or dispute these errors. To mitigate this issue, participants emphasized that beyond establishing clear policies and guidelines, future versions of the CAT tool should help users, especially new ones, thoughtfully evaluate AI-suggested tags and select accurate Depicts statements. As Paul explained:

\begin{quote}
\textit{"...or maybe even a tutorial that starts by asking people to tag images where the tags are already known and giving them bad examples, saying, ‘No, you shouldn't have picked that.’ And the reason you shouldn't is because... I've seen this in other crowdsourcing tools. For example, I worked with a tool using NASA images of Mars to identify geological features. You had to complete a training exercise where it asked, ‘Which of these things are in this image?’ If you picked the wrong one, it explained why it was incorrect and provided key points to learn from. Once you reached a level where your answers were consistently correct, it allowed you to move on to new tasks."} - Paul
\end{quote}

Paul’s proposal highlights a broader need for a structured onboarding experience that helps contributors understand what Depicts statements mean, learn from incorrect tag selections, and build confidence in applying ontology-linked structured metadata accurately over time.

\subsection{Lack of Integration between Categories and CAT}

Categories have been a central organizational tool in Commons since its inception. They allow users to group multimedia files into broad classifications. For many years, the category system provided a lightweight yet effective means of browsing the repository and locating content through keyword search. However, as Commons has grown to host millions of files, the limitations of this system have become increasingly apparent, not only to the Commons editing community, but also to the Wikimedia Foundation and researchers \cite{thornton2012tagging}. Our interview participants largely echoed this recognition: while categories have historically played an important role, they no longer meet the evolving needs of Commons. A shift toward structured metadata is seen as necessary to improve accessibility, enable more accurate knowledge representation, and improve search capabilities. As Paul explained:

\begin{quote}
\textit{“I think the end goal will be that we don't need to use categories, or we are much less reliant on them. Because structured data allows you to say more things and to do it in a more machine readable and a more comprehensive way, and with fewer of the issues that arise from categories being misapplied or the limitations of categories, like circular categories and category inheritance being misleading …”} – Paul
\end{quote}

Despite recognizing the limitations of categories, participants strongly opposed the current CAT approach of creating a parallel structured data system, which they saw as dismissive of the collaborative knowledge embedded in the category system. They argued that instead of discarding the category system and "reinventing the wheel," structured data should be integrated with it. As Jack said:

\begin{quote}
\textit{“And I still think they made a terrible mistake by not thinking at all of how structured data could integrate with categorization, instead of being something sitting side by side for a parallel purpose… And there was, as I say, no respect for looking at what was already built by the community… and [it] was an enormous amount of effort compared to what it would have been to refine what was already there, if they'd had any respect for it.”} – Jack
\end{quote}

To the participants, categories are not just navigational aids; they represent a repository of collective knowledge accumulated through around twenty years of Commons community-driven work in organizing, curating, and annotating multimedia content. These human-generated category tags capture semantic and contextual knowledge that CAT suggestions, AI-generated structured descriptive labels, cannot fully reflect. Harris illustrated this distinction:

\begin{quote}
\textit{“Yes, the depicts statements are more what is… visible. And the categories… I think they are more based on the metadata about what is behind a picture. So let's take this picture of 1995. It's a historical moment. I described all the names of these men because I know them. So this is… information I can give… not only in the description, but also in the categories. But depicts statements are more on the surface, on… what can be seen. Is it outside or in a room?”} – Harris
\end{quote}

Such knowledge, often encoded in categories, can support structured data efforts beyond the surface level of visual depiction (the Depicts statement). They can inform other structured data properties related to historical context, artistic medium, and provenance. As Paul noted:

\begin{quote}
\textit{“Categories at one level can be the same as depicts so that drawing is in the category Aston Lower Grounds, which is what is in the depicts statement… But the categories also have 1880s… in Birmingham, which is that area is now part of Birmingham. At the time it was not part of Birmingham. So putting Birmingham in a depict statement is not appropriate. The category has the fact that it was in a book. It has the book that it was taken from… but that is not what it depicts. It doesn't depict the book. You can't look at that and say, Oh yes, I can see a book, but the category tells you that it was in that book. The category tells you that it is a drawing and not a photograph or a painting, and that is a property of the image, but it is not what the image depicts. The category may tell you who the artist (who drew it) was, but that is not depicts… Other structured metadata can tell you that, but not the depict statement.”} – Paul
\end{quote}

Therefore, participants suggested that SDC and future CAT tools should build upon, rather than discard, the legacy of Commons categories by integrating knowledge from existing category tags into structured metadata. For example, as Jack explained:

\begin{quote}
\textit{"I would hope, if something has that already there with a category, and we're trying to do things on an automated basis, it could know, oh, this is in Woodland Park... Take a hunk of text and do some analysis on it with AI."} - Jack
\end{quote}

Jack suggested that future CAT tools could analyze existing category tags to extract structured property–value statements, map them to Wikidata identifiers, and present them as suggestions to help editors transform the knowledge embedded in categories into structured metadata.

\subsection{Ill-specified AI/ML Tasks}

The CAT tool applies a generic AI/ML algorithm (Google Cloud Vision API) to label all multimedia files in Commons and suggest tags. Participants expressed concerns about this ill-specified task—applying a one-size-fits-all AI/ML model to the entire repository for unsupervised discovery—because it often generates overly broad suggestions that are not useful for editors and users. As Sam explained, this generic AI/ML application is ineffective because AI/ML algorithms perform best when trained for clearly specified tasks. He elaborated:

\begin{quote}
\textit{“This is not mission oriented machine learning. This is like theoretical broad, whoever user version of machine learning… I think this descent into, like, over generic content, right? Um, what we're finding is, and all applications of AI really, like, the good uses are very specific users with a specific training set, right? Like…I want to find galaxies and, like, 10 million images from the Hubble Space Telescope, right? Like, it's really that narrow, the really good uses. And the more and more generic it becomes, the more and more bland it becomes… Specificity generates usefulness from them.”} - Sam
\end{quote}

In interviews, participants provided examples of clearly specified tasks that AI/ML models could handle effectively within Commons. Kirk, for example, highlighted the potential for machines to identify and label SVG flag images based on specific letter inscriptions. He explained:

\begin{quote}
\textit{“… if I had an SVG flag of Washington…and it says, I can't remember if the state of Washington has the word Washington written on it, but if it's encoded in actual text, then a machine can add in categories about inscription by letter perfectly…That is one example of things I added to these flags, is when flags have inscriptions on them, I'll say it has the letter W, it has the letter A, it has letter S. That'll be perfect for a machine to do. But that's the kind of niche categorization system isn't it? … So even for like, the most narrow, simple tasks, they're only good as long as we put proper semantic information in the files, like I mentioned, like if you have SVG and you mark it up correctly, and you tell it what it needs to know. It can interpolate and it can do things with the data. But if you're taking just general photographs of stuff and hoping that machines can figure out what's in them…They can't be trusted to discriminately, just add tags and categories and depictions and structured data.”} - Kirk
\end{quote}

This example illustrates how AI can assist with clearly specified, repetitive tasks that are not subjective, such as identifying specific features in flags, which would allow human contributors to focus on more complex work. Sam also proposed that AI could assist with high-volume tasks, such as categorizing large sets of similar images, including identifying the color of fireworks. He explained:

\begin{quote}
\textit{“Now, the ideal method for deploying something like this would be to give a community member control like, I want color tags for these 30,000 there are like 30 or 40,000 images of fireworks on Commons, right? That's like a ridiculous number of fireworks. And I'm sorry, you don't really need a human to tell you what color they are, you could just run Google visions tool and be like, here's your color range. The color parallel for fireworks is, like, 10, right? And like, tell me which one of these, each of them is. And like, 95\% of them would be accurate, right? And then we wouldn't have to worry about depicting the fireworks by color with humans, we could focus on the more interesting human tasks, right? And so I think this is the problem that happened, is like they weren't looking for backlogs, or they weren't empowering the community to choose which backlogs were handling, right?”} - Sam
\end{quote}

Sam’s example also highlights the importance of aligning AI/ML tools with community needs and priorities. When AI is used to address clearly specified tasks that the Commons community values—such as tagging the colors of fireworks—it can significantly reduce backlogs and improve efficiency. Moving forward, participants suggested the Wikimedia Foundation help identify and define clearly specified AI/ML tasks that the Commons community values. This can be done by fostering collaboration within Commons communities focused on specific domains to effectively identify and define the AI/ML tasks required to support their work, as Sam explained:

\begin{quote}
\textit{"The taxon group...has a very clear use case. And there's like, train machine learning models and like, they don't want the vag bird tags, like, those are not useful for anyone. And so I think there's like, you could subdivide it into domains pretty easily"} - Sam
\end{quote}

Additionally, participants advocated for the selection of specialized AI/ML algorithms tailored to clearly defined tasks in specific content domains, rather than using generic tools like Google Cloud Vision. Ron provided examples of algorithms for plant and flower identification, which have demonstrated expert-level accuracy:

\begin{quote}
\textit{“...the very good apps for plant and flower identification that already exist like PlantNet or Flora Incognita. These apps are so advanced that you can identify plants in the field simply by taking a smartphone image and running the app. In almost all cases, the identification is accurate, really expert level."} – Ron
\end{quote}

In conclusion, participants suggested that AI/ML tools are most effective on Commons when applied to clearly specified tasks that align with community priorities. In contrast, using generic AI/ML models for ill-specified tasks—such as broadly labeling diverse media files—produces inconsistent results and offers limited value. For future CAT tool design, participants called for a shift toward the collaborative identification of valued, specific use cases and the adoption of specialized models tailored to each case.

\subsection{Lack of Support for Collaborative Decision-Making and Evaluation}

We found that tagging images on Commons is a collaborative, iterative process. Individual contributors often struggle to assign the most accurate and appropriate Depicts statements on their own and instead rely on community input to review and refine them over time. However, the current design of CAT suggestions does not adequately support this collaborative reality. This is not a conflict but a contrast between how Commons contributors approach tagging and what the CAT tool currently supports.

In interviews, participants shared that they often struggled with the accuracy, relevance, and usefulness of the Depicts statement they assigned. This challenge became especially apparent when editors were uncertain about how others might interpret or use a given image. In response, many adopted a collaborative approach: they began with broad or generic Depicts statement with an expectation that more knowledgeable editors would later refine or correct them. As Simon explained when tagging an unfamiliar item of clothing:

\begin{quote}
\textit{"If I'm 100\% sure it's Salwar Kameez, I would tag it Salwar Kameez. But if I'm not sure, I would tag it just as cloth and hopefully someone who is very knowledgeable about Indian fashion would come and say, 'Okay, this is actually this kind of dress, this kind of cloth or something.' They would get it right, whereas if I do it, maybe I would get it wrong. So in such cases, it can be good to not take the risk."} – Simon
\end{quote}

This strategy reflects a common practice: using broad, "safe" Depicts statements like “cloth” to avoid misclassification, while trusting that the community will help improve accuracy over time. Another participant further explained the value of general Depicts statements as an entry point for collaboration. Even vague or general Depicts statements provide a foundation that others can build on:

\begin{quote}
\textit{"And I think that even something very broad is still better than nothing. Because people can then, like a specialist, can then come and, for instance, if I tag building, some other people could come, like specialists of buildings, could come and ventilating the proper depictions, such as, I don't know, castle or church and further down... I think it's a good first step to properly classify the picture, because if the picture has no depictions or categories at all, nobody would ever find it, except if it has a good name, but it's much easier to find and refine pictures that have a base category or base depiction already."} – Simon
\end{quote}

Some participants even admitted to using Depicts statements they knew were inaccurate, intentionally, in the hope that someone more familiar with the subject would later correct them. For example, Simon described tagging a kindergarten with a city name due to both time constraints and the absence of a relevant Wikidata item:

\begin{quote}
\textit{"There was no Wikidata item for this particular kindergarten... So I think I tagged with the city <name>... And I thought, okay, people who are in it, they will further refine this, so I don't need to do it, they will do it."} – Simon
\end{quote}

These examples show that tagging is often treated as an ongoing, shared task rather than a one-time decision by a single contributor. Editors act with the expectation that their work will be reviewed and improved by others. This collaborative process also extends to evaluating the usefulness and appropriateness of Depicts statements. Editors frequently reported difficulty judging the relevance of a Depicts statement on their own, since they cannot anticipate all the ways others might use or search for an image. Joel, for example, reflected on how the same image might serve multiple—even unexpected—purposes:

\begin{quote}
\textit{"I don't know what I would be searching for, you know, I guess maybe, I don't know, depicting that section of the shoreline? … The main subject, the main subject here will be cliff diving, you know. I think I used this particular image in the article related to the mathematical concept of parabolas. So you know, it's just, different usage."} – Joel
\end{quote}

Joel’s reflection shows how the potential uses of an image, for sports, geography, mathematics, and more, are diverse and difficult for any one editor to fully anticipate. This further highlights the need for collaborative sense-making and reasoning in both assigning and evaluating Depicts statements.

Taken together, these examples show that tagging images on Commons is often a collaborative, iterative effort. Editors rely on broad or generic Depicts statements when uncertain, and these initial statements serve as placeholders that are later refined or corrected by others. The subjective nature of relevance and the varied use cases of images make tagging a process that benefits from, and often requires, ongoing collaborative decision-making and reasoning. While current contributor practices require collaboration, CAT suggestions assume that the uploader or tagger will make the final decision alone. The system provides no mechanisms for shared review, discussion, justification, or iterative refinement of suggested Depicts statements. It surfaces suggestions to individuals but does not support contributors in discussing why a tag is selected, comparing it with alternative interpretations, or inviting structured input from others before a tag is finalized. This limits CAT’s ability to function as a tool for collaborative validation, even though the community often treats initial Depicts statements as support for coordination rather than a terminal labeling decision.

\subsection{Disconnect Between CAT and Commons’ Search Functionality}

One of the central goals of the CAT tool was to improve the searchability of Commons content through the addition of structured data. However, many participants reportedly stopped contributing to CAT and Structured Data on Commons more broadly because they felt the current implementation did not meaningfully improve search functionality on Commons.

Currently, the CAT tool helps connect Commons' text-based search to Wikidata by adding Depicts statements. In an effort to support both general and specific queries, contributors often include both broad terms (e.g., “man” or “woman”) and highly specific ones (e.g., “Vincenzo de Cotiis”) in their Depicts statements via CAT. However, this practice runs counter to established best practices on Commons and Wikidata, which recommend selecting only the most specific, non-redundant term and avoiding repeating information already implied by a hierarchy. Participants noted that this approach has inadvertently created a parallel folksonomy resembling Commons' older category system. As one participant explained:

\begin{quote}
\textit{"They just ended up with a second, parallel folksonomy. They did not do the hard work of working out how you model data in a given area before releasing to the general use in that area. They just built another folksonomy with a bunch of people who, you know, doing it by losing texts over time. Yeah. And so it's no better or worse. It’s, you know, in Wikidata itself, because they have that additional tool of expressing their relations, [it] is ontologically a bit better than Commons. But SDC is not."} - Jack
\end{quote}

This over-reliance on exhaustive term inclusion has yielded limited gains compared to the traditional category system. The deeper issue lies not in CAT itself, but in how queries are processed by the search system. While CAT and Structured Data on Commons have the potential to enhance search, by leveraging semantic relationships and hierarchical reasoning in Wikidata, these benefits remain largely unrealized in current search implementations. One participant described this missed opportunity:

\begin{quote}
\textit{“Well, you're obviously including some keywords, and search engines can be taught to understand the structured data, so that, for instance, if you search for Beagle, you can indicate to the search engine whether you're looking for dogs or the ship that Darwin sailed on—his famous boat the HMS Beagle, for example. So it's disambiguating similar terms, but it's also understanding that if you search for dog, then you might want Beagle. You don't just want things that are necessarily labeled as dog. So that's how structured data can help with the searching side, rather than the querying side. Now, you have to be more clever than that with your search query. So you can search for things that have particular structured data statements, like depicting a beagle dog, but you have to know how to do that. You have to be able to understand a little bit of the markup code to do that, whereas eventually, I hope it will be written into the search engine. So if people just type in plain language, then it will find things using that mechanism.”} - Paul
\end{quote}

Because the current system places the burden on contributors without delivering an improved search experience in return, many participants have disengaged not only from CAT but also from Structured Data on Commons more broadly. They viewed the process as ineffective and contributing to the pollution of Depicts statements and the re-creation of a folksonomy similar to the old category system. To address this challenge, participants suggested that, in addition to improving CAT, Commons should design and deploy a search mechanism that fully leverages Wikidata’s semantic relationships and aligns with best tagging practices across the Wikimedia ecosystem.

\section{Discussion}

In this paper, we investigated Wikimedia Commons contributors’ lived experiences with the Computer-Aided Tagging (CAT) tool using two complementary data sources: 595 user comments from 11 wiki pages and 16 in-depth interviews. Our analysis revealed seven key issues that contributed to CAT’s mixed reception and eventual deactivation: (1) misalignment in the perspectives of Structured Data on Commons, (2) unclear definitions of the Depicts statement, (3) difficulties applying Depicts through CAT, (4) lack of integration between categories and CAT, (5) an ill-specified AI/ML task, (6) limited support for collaborative evaluation, and (7) a disconnect between CAT and Commons’ search functionality. Beyond identifying these challenges, we offered community-informed suggestions for improvement. In the remainder of this section, we discuss the broader implications of these findings for designing human–AI collaboration on Commons and for developing AI-assisted tools that support open knowledge work.

\subsection{Designing for Human–AI Collaboration on Commons}

This subsection outlines key considerations for designing AI-assisted tools that enable effective human–AI collaboration on Wikimedia Commons. Based on our findings, we emphasize the importance of fostering consensus-building, adopting participatory AI/ML approaches, leveraging legacy systems, and addressing linguistic and cultural gaps.

\subsubsection{Fostering Consensus-Building}

Our findings highlight the important role of consensus-building in the effective design and deployment of AI-assisted tools on Commons. While tools like CAT are developed with good intentions, they can unintentionally disrupt established workflows and community values when consensus—around project goals, task definitions, and best practices—is lacking. In the case of CAT, the absence of shared norms regarding the goals of the Structured Data on Commons project and the CAT tool, the appropriate scope of AI-assisted tasks, and what constitutes a valid Depicts statement made it difficult for contributors to engage with or trust the system’s suggestions.

In the CSCW and HCI literature, consensus building has been extensively studied in the context of Wikipedia \cite{kriplean2007community,im2018deliberation}. Researchers have developed a variety of sociotechnical systems to support Wikipedia's consensus building practices: tools that help identify emerging conflicts \cite{kittur2007he}, systems that promote mutual understanding through structured summarization \cite{kriplean2012you}, and mechanisms for collaboratively producing discussion summaries \cite{zhang2017wikum}. Wikipedia also benefits from formal governance structures that identify consensus as a central principle supported by detailed policies\footnote{\url{https://en.wikipedia.org/wiki/Wikipedia:Consensus}} and community-authored guidance\footnote{\url{https://en.wikipedia.org/wiki/Wikipedia:Method_for_consensus_building}} on how to achieve it.

In contrast, Wikimedia Commons has received comparatively little research attention and lacks parallel governance structures. There are no formal policies, established guidelines, or tools to support Commons contributors in building consensus. This difference is rooted in the distinct nature of the two projects. Wikipedia is a collaborative text-based encyclopedia, where contributors must synthesize diverse perspectives into coherent, neutral articles, which naturally necessitate deliberation and consensus. Commons, on the other hand, is a multimedia repository where contributions—such as uploading, adding metadata and categorizing files—are typically carried out independently, with minimal coordination or conflict resolution. 

This structural independence is well documented in prior research \cite{yu2022unpacking, yu2023you}, which notes that Commons workflows often operate with low levels of coordination and without a strong expectation of consensus. Our findings support this view. Participants described using their own interpretations when applying Depicts statements, often in the absence of any shared standards. In some cases, small subgroups had developed informal alignment, but there was little indication of platform-wide consensus. While this openness supports diverse participation and flexible contributions, it also creates challenges when introducing AI tools that rely on common ground to function effectively. 

These findings suggest that the challenges faced by CAT were not only technical but also socio-organizational. Building successful tools to support human–AI collaboration on Commons requires more than accurate algorithms; it requires attention to the social infrastructure needed to foster alignment. Rather than assuming consensus exists, future tool design and development must actively support its emergence. Future research could explore sociotechnical approaches to support consensus-building in Commons’ environment, which is even more decentralized than Wikipedia.

\subsubsection{Adopting Participatory AI/ML Approaches}

Our findings show that a major reason behind the CAT tool's low adoption, usage, and satisfaction among Commons editors is its usage of a generic AI model for an ill-specified task—using the Google Cloud Vision API to broadly label the entire Commons repository for unsupervised discovery. This approach often results in overly broad or irrelevant suggestions that offer limited value to Commons editors and disrupt their existing workflows. Participants expressed a preference for more narrowly scoped models trained to address specific, well-defined tasks aligned with community priorities. These findings highlight that designing AI/ML tools to support work on Commons requires early and ongoing engagement with the communities who use and sustain the platform. Tools must be aligned with the needs, values, and everyday practices of contributors to be useful and accepted.

In other open knowledge environments where contributions are similarly decentralized and community-driven—such as Wikipedia—participatory approaches (e.g., ORES \cite{halfaker2020ores, smith2020keeping}, LiftWing\footnote{\url{https://wikitech.wikimedia.org/wiki/Machine_Learning/LiftWing}}) to AI/ML have been introduced to support this need. Unlike CAT’s reliance on a single “best” classifier embedded by default, participatory AI/ML frameworks allow for the coexistence of multiple, potentially contradictory classifiers. The outcome of participatory AI/ML is not a singular model or tool, but a public, collaborative process of training, auditing, reinterpreting, appropriating, contesting, and negotiating both the models themselves and their deployment in content moderation interfaces.

Applying such an approach to Commons presents unique challenges. One central difficulty lies in identifying cohesive communities of Commons editors who share domain knowledge, editorial goals, and workflows. In Wikipedia, language versions often serve as natural community boundaries which enable collaborative development and deployment of participatory AI/ML tools, as suggested in prior research \cite{halfaker2020ores}. In contrast, Commons is a truly multilingual platform that lacks comparable linguistic segmentation. While many Commons contributors originate from specific language editions of Wikipedia—initially using Commons to manage media files for their home Wikipedias—only a small portion identify Commons as their primary project \cite{yu2022unpacking}. As such, strong community ties within Commons itself are relatively rare. Most contributors work independently on tasks that do not require sustained collaboration or community building \cite{yu2023you}.

This more decentralized and individualistic structure complicates efforts to initiate participatory AI/ML processes on Commons. As our participants suggested, a promising starting point may be to map content domains (e.g., biological species) and identify opportunities for collaboration within and across these domains. However, given the cross-linguistic nature of Commons, such efforts will need to address significant linguistic and cultural barriers. In domains where contributors from different language versions of Wikipedia converge, effective participatory AI/ML will require tools and practices that facilitate multilingual collaboration and interpretation.

\subsubsection{Leveraging Legacy Systems}

Wikimedia Commons was launched in 2004. It has served internet users for over two decades. During this time, internet technologies have evolved rapidly, yet Commons has continued to rely on longstanding community-driven methods for accumulating and organizing collaborative knowledge. While these methods—including the traditional category system—may not fully meet the demands of Commons’ evolving size and scope, they remain an important legacy. Rather than being discarded, these systems should be leveraged and integrated into the design of new AI-assisted tools.

In our study, participants acknowledged the limitations of Commons’ category system, particularly in supporting modern, semantic search. However, they strongly opposed the CAT tool’s approach of bypassing this legacy system to introduce an entirely parallel tagging infrastructure. Participants viewed this as a rejection of the community’s accumulated practices, norms, contributions and existing knowledge.

This tension between structured tagging and unstructured user-generated tags has been discussed in prior research in image processing and computer vision. For example, social media platforms such as Flickr and Zooomr have long allowed users to manually tag their images with freely chosen labels. These tags support image indexing and search, but due to their grassroots nature, they often diverge from the actual visual content of the images \cite{liu2011content}. Research \cite{kennedy2006search} found that when a tag appears in a Flickr image, there is only about a 50\% chance that it accurately reflects the content. Similarly, studies \cite{bar2008structured,sigurbjornsson2008flickr} demonstrated that unstructured tagging allows for a wider range of tags, and that many user-assigned tags describe contextual information—such as location, time, or sentiment—rather than what is visually depicted.

This distinction is important because unstructured, user-generated category tags—though often inconsistent—contain valuable contextual information that structured Depicts statements alone cannot fully capture. These tags frequently reflect metadata such as location of creation, event, or creator, which can enrich the semantic understanding of media files. Rather than treating these human-generated category tags as outdated or irrelevant, future AI tools should use them as a complementary source of knowledge.

One promising strategy for future CAT design is tag processing \cite{liu2011content,wang2012assistive}, an approach that automatically analyzes and refines user-provided category tags to suggest more accurate and semantically aligned structured metadata statements. By applying tag processing techniques, developers can bridge the gap between Commons’ rich but informal categorizing practices and the platform’s emerging structured data infrastructure.

\subsubsection{Incorporating Multilingual Image Descriptions}

One major goal of SDC and the CAT tool is to reduce multilingual barriers to content access and contribution on Wikimedia Commons, a media repository that supports over 300 Wikipedia language editions and users worldwide. Contributors can upload files and provide descriptions in any language, which makes traditional, unstructured Commons metadata multilingual by design. While this decentralized model diversified cultural knowledge coverage, prior research \cite{yu2022unpacking} has shown that English-dominant descriptions created accessibility gaps for contributors and users without English proficiency. This barrier is not one-sided, it also affects English-speaking contributors, who often struggle to locate materials about non-English-speaking regions, because relevant descriptions and curation work are frequently produced in the primary language of the source community \cite{yu2022unpacking}.

To mitigate this access gap, CAT ignored existing multilingual image descriptions, analyzed only image pixels at upload, and mapped suggested labels to language-agnostic Wikidata Q-IDs. This design enabled contributors from different language communities to evaluate a shared set of suggestions using equivalent information. However, as our findings show, participants noted that this design choice led to a shift toward more generic image tagging, which many feared would undermine Commons’ educational purpose. This concern is tied to a broader issue: major contemporary vision models, including Google Cloud Vision, are trained on linguistically and culturally skewed, web-scale datasets, and their tag suggestions risk flattening culturally specific concepts into generalized, Western-default labels \cite{liu2025culturevlm, jeong2025culture, yun2024cic, birhane2023hate, luccioni2023stable}.

To address this gap, future CAT tools should incorporate, rather than bypass, Commons’ multilingual textual metadata when generating tag suggestions. Specifically, future designs should combine pixel-based vision inference with language-based processing of unstructured multilingual descriptions. This would enable systems to propose Depicts statements that are semantically specific and culturally contextualized, instead of relying on generalized Western-default labels.

\subsection{Designing AI-Assisted Tools to Support Open Knowledge Work}

This subsection discusses key considerations for designing AI-assisted tools for open knowledge work, with a focus on supporting ambiguous and multimodal tasks, and expanding the scope of human–AI collaboration beyond moderation.

\subsubsection{Supporting Ambiguous and Multimodal Tasks}

This study expands existing knowledge by showing that the traditional approach to AI design for supporting open knowledge work \cite{adler2011wikipedia}—where the goal is often to eliminate ambiguity—does not apply in all open knowledge contexts.

Our findings show that designing AI-assisted image tagging in open knowledge systems is challenging due to the subjective nature of the task—tagging images is not a straightforward process like making predictions on textual data. Unlike textual data, which tends to follow more rigid rules and structures, images are multimodal and require interpretation informed by culture, context, and individual perspective. As we’ve seen in this study, both AI systems and human contributors struggled with ambiguity when assigning Depicts statements.

In the context of image tagging, AI systems should be designed not to eliminate ambiguity but to navigate and support it. Rather than providing definitive labels, AI tools should present suggestions that reflect the complexity and uncertainty inherent in visual data. An effective AI design should make these complexities and uncertainties visible to contributors, communicate the confidence levels of AI-generated suggestions, and provide clear contextual information about how AI arrives at its suggestions and what factors influence its decisions. This would enable human contributors to make more informed choices about whether to accept or refine AI-generated labels.

\subsubsection{Expanding the Scope of Human–AI Collaboration Beyond Moderation}

The introduction of AI tools like the CAT tool for content creation marks a significant shift in how humans and AI collaborate within open knowledge systems. Historically, AI tools on platforms like English Wikipedia (e.g., ORES \cite{halfaker2020ores}, SuggestBot \cite{cosley2007suggestbot}) have focused on indirect tasks such as moderation and coordination, rather than direct content creation. In contrast, CAT’s role in generating content—specifically, generating Depicts statements for image files—expands AI’s function from that of a supportive tool to that of a co-author.

This shift introduces new challenges for technology design, particularly in understanding how human contributors perceive AI authorship. As our findings show, CAT’s contributions to the content layer blur the traditional boundary between human editors and machine-generated contributions. Contributors expressed concerns about the AI’s limited ability to interpret the context and meaning of visual data. Accordingly, they suggested that human–AI collaboration in CAT should be designed as a co-creative process, in which the AI acts as a partner that offers suggestions to inform decision-making without overriding human judgment.

Therefore, we suggest that future research adopts a broader perspective on designing for human–AI collaboration in open knowledge work—one that investigates AI’s growing potential to reshape the dynamics of authorship, responsibility, and editorial control in collaborative content creation.

\subsection{Limitations}

This study has several limitations. First, our findings are based on qualitative data from 160 users and 16 in-depth interviews. While these sources provide rich insights, they may not fully capture the diversity of the broader Commons community. Contributors who engaged with the CAT tool and participated in discussions may differ significantly from those who chose not to engage.

Second, we were unable to reach users who had negative experiences with CAT and subsequently disengaged from Commons. As a result, certain critical perspectives—particularly those shaped by frustration or dissatisfaction—may be underrepresented in our analysis.

Third, most of our interview participants were experienced, active users. Consequently, the study may reflect the perspectives of veteran contributors more strongly than those of newcomers or occasional participants.

\section{Conclusion}
In this paper, we present a qualitative analysis of 595 user comments and 16 interviews exploring Wikimedia Commons contributors' lived experiences with the CAT tool. Our work provides an empirical understanding of seven key challenges that shaped CAT’s mixed reception and eventual deactivation, along with community-informed suggestions for improving the tool. We expand the understanding of human–AI collaboration beyond Anglophone, text-centric, corporate platforms. This study highlights the exciting potential for future research at the intersection of AI, image tagging, and decentralized, multilingual communities.

\bibliographystyle{ACM-Reference-Format}
\bibliography{sample-base}

\appendix

\section{Interview Protocol}
\label{app:script}

\subsection{Opening Script}

Hi! I’m <First Author>. Today, I’d like to talk with you about your experiences with the now-deactivated computer-aided tagging (CAT) tool on Commons. In this interview, I have a set of questions to guide our conversation, but I may also ask follow-up or clarification questions as we go.

Please know that your participation in this study is entirely voluntary. If at any time you’d like to stop the interview, just let me know, and we’ll stop. Also, if there’s a specific question you’d prefer not to answer, just let me know, and I’ll move on to the next one.

With your permission, I’d like to audio record the interview. Only my advisor, <Last Author>, and I will have access to the recordings. We will transcribe the interviews and erase the recordings afterward.

In case of poor network connection during the interview:

\begin{itemize}
    \item If my connection drops, please stay online and wait for at least five minutes for me to rejoin.
    \item If your connection drops, please try to rejoin using the same Meeting ID. I’ll wait for you for up to five minutes.
    \item If we cannot resume the interview within five minutes, I’ll contact you to reschedule and send the details by email.
\end{itemize}

We make a good-faith effort to maintain the confidentiality of this interview. We won’t disclose that you participated in this research, and we’ll do our best to anonymize the data. However, I should let you know that sometimes Wikipedians can identify individuals or situations, even though we work to anonymize everything. Do you have any questions before we begin?

<Start Recording>

Just for the record, is it okay for us to audio record your interview?

\subsection{Phase 1: Introduction/Warm-Up}
Q1: How long have you been editing Wikimedia Commons?\\
Q2: What types of work do you currently do in Wikimedia Commons?\\
Q3: In your opinion, what is the goal of structured data for Commons content?
\subsection{Phase 2: Tasks and Activities}
Now, I would like to show you some of the images you uploaded to Commons and ask you some questions about them. I’m going to share my screen so that you can see what I see.
\begin{itemize}
    \item Contribution 1 (Upload image)
    \item Contribution 2 (Upload image)
    \item Contribution 3 (Upload image)
\end{itemize}
For each contribution, ask:\\
Q4: What is the story behind this contribution?\\
Q5: What do you think the image Depicts, and why?\\
Now I’ll show you a list of suggested tags for this image generated by the CAT tool.
\begin{itemize}
    \item Tag List 1:
    \item Tag List 2:
    \item Tag List 3:
\end{itemize}
Q6: Could you explain the value of these tags?
\begin{itemize}
    \item Follow-up: Do you think these tags make the image easier to find in Commons?
\end{itemize}
Q7: How would you tag this image yourself, and why?
\begin{itemize}
    \item Follow-up: These tags differ from what you said the image Depicts. Can you explain why?
\end{itemize}
Q8: When tagging an image like this, are there any metadata (e.g., categories, file name, file description) you find important to consider?
\subsection{Phase 3: CAT Tool}
Q9: What do you remember about the CAT tool?\\
Q10: How do you think the tool impacted your work in Commons?\\
Q11: What does the category system in Commons do?\\
Q12: What does the Depicts statement in Commons do?\\
Q13: Does the CAT tool handle one of these functions: Categorization or Depicts?\\
Q14: In your opinion, what is the goal of structured data for Commons content?\\
Q15: How do you think the CAT tool served that goal?\\
Q16: In your opinion, how does the CAT tool support other Wikimedia projects, such as Wikidata?

\subsection{Phase 4: AI Bias}
Q17: What are the risks associated with using AI to tag images?\\
Q18: How do you think we can mitigate these risks in Commons?

\subsection{Closing Script}
Do you have any questions for me?\\
We acknowledge the time and effort of our participants by sending a \$25 Amazon gift card. Do you have a preferred email address that you would like us to use to send the gift card?
\begin{itemize}
    \item We can also send Visa gift cards or make a donation in your name to one of three charitable organizations: the Wikimedia Foundation, Creative Commons, or the Internet Archive.
\end{itemize}
Thank you for taking the time to talk with me about Wikimedia Commons. Is there anything else you would like to add?

\end{document}